\documentclass[runningheads]{llncs}
\usepackage[T1]{fontenc}
\usepackage{tikz}
\usepackage{graphicx}
\begin{document}
\title{Applying Machine Learning for characterizing social networks Agent-based models}
%
%
\author{Haoyuan Li\orcidID{0009-0003-7608-1396} \and
Lidia Conde Matos\orcidID{0009-0001-8240-154X} \and
Eduardo César Galobardes\orcidID{0000-0002-9729-8557}\and
Anna Sikora \orcidID{0000-0003-0090-4109}}
\authorrunning{Haoyuan et al.}
%
\institute{Universitat Aut\`onoma de Barcelona, Cerdanyola 08193, Spain 
}
\maketitle              
\begin{abstract}
Nowadays, social media networks are increasingly significant to our lives, the imperative to study social media networks becomes more and more essential. With billions of users across platforms and constant updates, the complexity of modeling social networks is immense. Agent-based modeling (ABM) is widely employed to study social networks community, allowing us to define individual behaviors and simulate system-level evolution. It can be a powerful tool to test how the algorithms affect users behavior. To fully leverage agent-based models, superior data processing and storage capabilities are essential. High Performance Computing (HPC) presents an optimal solution, adept at managing complex computations and analysis, particularly for voluminous or iteration-intensive tasks. We utilize Machine Learning (ML) methods to analyze social media users due to their ability to efficiently process vast amounts of data and derive insights that aid in understanding user behaviors, preferences, and trends. Therefore, our proposal involves ML to characterize user attributes and to develop a general user model for ABM simulation of in social networks on HPC systems.

\end{abstract}

\section{Introduction}

Our long term goal is to define a general agent-based model for social media networks and investigate how they can be used to better understand the dynamics of these networks. Given the expanding influence of social networks, understanding their dynamics has become crucial. ABM allows researchers to simulate the behavior of individuals within a network and investigate how their interactions give rise to emergent properties of the network as a whole.

We aim to implement the ABM in a HPC platform that could realize the full potential of ABMS. Due to the scale and complexity of agents, implementing a parallel social network simulator on HPC is essential. We intend to define the model first and then explore how ABM can simulate user behaviors, information spread, and analyze their dynamics. Through our research, we hope to provide insights into the underlying mechanisms that govern social media networks. 

To accomplish our goal, we first need to model social network users. However, this is challenging due to the lack of labeled datasets. We also aim to understand the statistical distribution of user attributes within each category to create synthetic datasets. In this work we present our methodology for user classification using ML techniques and the preliminary results obtained by using Twitter datasets.

\section{Research Methodology}

\begin{figure*}[h]
    \centering
    \begin{minipage}{1.1\linewidth}
    {\footnotesize
      \includegraphics[width=12cm]{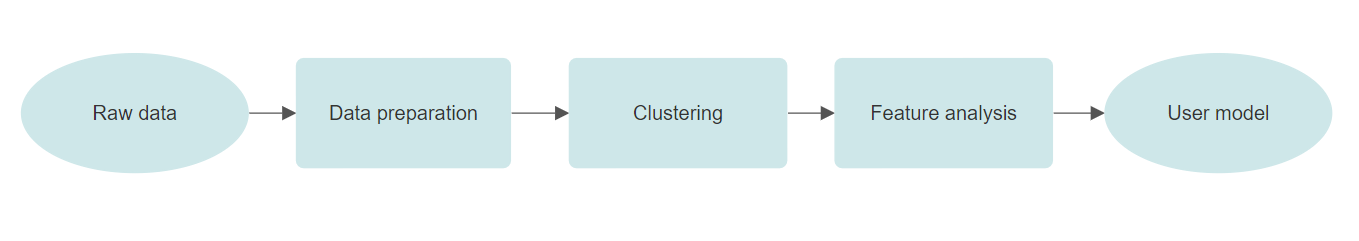}
   
    }
    \end{minipage}
    \caption{General steps of the proposed methodology}
    \label{flowchartl}
    \end{figure*}

We started from collecting social network users data and creating our own dataset which contains the attributes that can be analyzed for characterizing users. Then, with all the information gathered, we focus on the evaluation of various attributes by using Principal Component Analysis (PCA) \cite{P1}. Subsequently, we apply a ML method to classify users with these identified attributes, and then obtain the distribution patterns of each attribute in each group to generate a user model. The general steps of this work are illustrated in Fig.\ref{flowchartl}.

\vspace{-2\baselineskip}
\subsection{Data Preparation}

To build an effective dataset, ample data gathering and processing are essential. Many social media platforms provide APIs that allow for accessing to user-generated
text, images and videos, as well as to accompanying metadata. We selected the most distinctive attributes and transformed the raw data to include only these selected attributes.

A dataset typically encompasses both numerical and binary attributes, capturing diverse user characteristics and behaviors. Due to this diversity, it's vital to assess each attribute's usefulness in user classification. This evaluation aims to identify the most informative attributes for effective classification.

We have employed Principal Component Analysis (PCA) for attributes evaluation. As it can reduce the dimensionality of a dataset with numerous interrelated variables, preserving as much variation as possible. This is achieved by transforming to a new set of variables, the principal components (PCs), which are uncorrelated and which are ordered so that the first few retain most of the variation present in all of the original variables\cite{0}. To determine the number of principal components (PCs) to preserve in PCA, several common methods exist, including the Eigenvalue Criterion, Scree Plot, and consideration of Domain Knowledge. In this study, we will determine the component count by leveraging insights from prior research. To interpret the correlation of attributes usually means looking at the loadings in a PC and hence deciding which variables are important for that component \cite{1}. 

After performing PCA, we can assess all the attributes collected and determine which ones to retain for next steps. 
\subsection{Clustering using Machine Learning}

To do the clustering, we employed Machine Learning (ML), which offers the capability to handle high-dimensional and heterogeneous data, ensuring robust classification. ML methods can deal with various types of data, including numerical, categorical, and mixed attribute types. In ML, classification primarily involves supervised and unsupervised approaches. Supervised methods rely on labeled datasets for training, whereas our datasets lack. Hence, we opted for unsupervised methods to cluster the data. There are numerous options available for unsupervised methods such as K-means \cite{ML1}, K-prototypes \cite{ML4}, Hierarchical Clustering \cite{ML2} and BIRCH \cite{ML3}. We selected K-means because it is more computationally efficient, easy to implement and produces good results.

K-means is a popular unsupervised learning algorithm used for clustering which focuses on finding hidden patterns or structures in the data without the help of labels or previous answers \cite{K1}. The algorithm consists of two separate phases. First, it selects k centers randomly, where the value k is fixed in advance. Second, to take each data object to the nearest center\cite{K3}. Normally, the number of clusters is defined by using the elbow method, however in this study we chose a fixed number based on the distributions and prior analysis on how social network users are typically grouped. The clustering process finished when all the data objects are included in some clusters.

\subsection{Feature Analysis}

The last step consists in analyzing the statistical distribution of attributes within each group. To achieve this, we need to identify these distributions. The methodology involves an examination of attribute distributions within each cluster obtained through the K-means algorithm. We initially analyzed histograms to gain insights into probable distributions associated with each attribute. Next, we conducted goodness-of-fit tests, using the Residual Sum of Squares (RSS) metric to compare observed frequencies with those expected from theoretical distributions. Finally, we assessed goodness-of-fit using probability plots. This allows us to generate new users mirroring real user characteristics' distributions and validate them by comparing distributions.

\section{Preliminary Results: Use Case}

In this section, we apply the methodology to characterize users using a real raw dataset of Twitter tweets.

\subsection{Step 1: Data Preparation}

The data used in this work consisted of a collection of JSON files grabbed from the general Twitter stream dated November 2022 \cite{URL} which contains tweets for more than 130000 users. We processed this data to obtain the users' attributes. The first column of Table \ref{loading} shows all the attributes we chose.

Then, to effectively classify users based on these attributes we used PCA. Table \ref{loading} displays attribute loadings in the first components. We can see that most of the numerical attributes have higher loadings than the binary ones indicating their greater influence on the principal component. Binary attributes are not the important part for classifying users. In the end we chose friends\_count, followers\_count, listed\_count, favourites\_count, statuses\_count for the clustering.


 \begin{table}[htb]
\caption{PCA loadings of all the attributes}
\begin{center}
 {\small
\begin{tabular}{|c||c|c|c|}\hline
Attributes & PC1 & PC2 & PC3   \\\hline\hline
location       & 5.313587e-09 & 1.467293e-07 & -5.952920e-07 \\\hline
protected       & -0.000000e+00 & -1.110223e-16 & -1.110223e-16  \\\hline
verified       & 3.946670e-08 & 3.000574e-08 & 5.879996e-08 \\\hline
followers\_count      & \textbf{9.999922e-01} & -3.192501e-03 & -1.511976e-03 \\\hline
friends\_count      & 8.453385e-04 & 4.239099e-03 & -1.111189e-02 \\\hline
listed\_count       & 1.544425e-03 & 3.381150e-04 & 2.707960e-04  \\\hline
favorites\_count       & -1.069482e-03 & \textbf{1.370760e-01} & \textbf{-9.904922e-01} \\\hline
status\_count      & 3.366800e-03 & \textbf{9.905463e-01} & \textbf{1.371111e-01} \\\hline
created\_at       & -3.165097e-07 & -3.738327e-06 & 1.040501e-05\\\hline
geo\_enabled       & 1.439318e-08 & 2.101619e-07 & -6.423289e-07  \\\hline
default\_profile       & -2.844280e-08 & -2.706425e-07 & 6.173020e-07 \\\hline
default\_profile\_image      & 0.000000e+00 & -0.000000e+00 & -0.000000e+00 \\\hline

\end{tabular}
\label{loading}
}
\end{center}
\end{table}

\subsection{Step 2: Clustering using Machine Learning Methods}

\begin{figure*}[h]
    \centering
    \begin{minipage}{1.1\linewidth}
    {\small 
      \includegraphics[width=12cm]{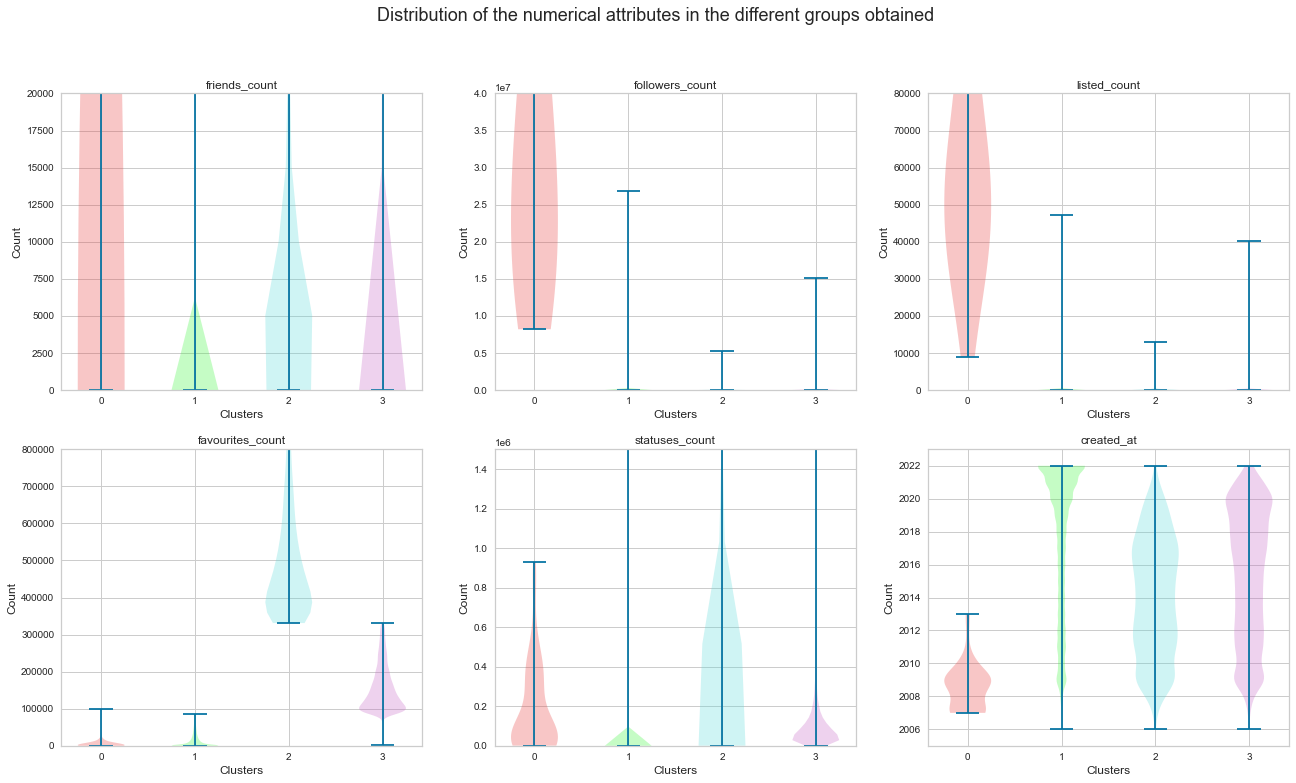}
   
    }
    \end{minipage}
    \caption{Distribution of the numerical attributes in the different clusters}
    \label{fig:kmeans_numerical}
    \end{figure*}

With all the data prepared, we can apply K-means in clustering dataset. In our case, we first employed elbow method to get the best cluster number which gave us K=7. Nevertheless, defining each of the 7 clusters obtained poses a challenge because some clusters exhibit a less distinct distribution of attributes. Consequently, based on the work \cite{10} and combined the analysis of our own case, we decided to determine the value of K by ourselves. We assumed K=4 after comparing the distribution.
Fig. \ref{fig:kmeans_numerical} shows the distribution of users in each cluster and Table \ref{tab:distribution} shows the percentage of users of each cluster.

 \begin{table}[htb]
\caption{Distribution of users per cluster obtained by the K-means algorithm.}
\begin{center}
{\small
\begin{tabular}{|c||c|c|}\hline
cluster & users & distribution   \\\hline\hline
0       & 61 & 0.000 \\\hline
1       & 115096 & 0.878  \\\hline
2       & 14158 & 0.108 \\\hline
3      & 1774 & 0.014 \\\hline

\end{tabular}
\label{tab:distribution}
}
\end{center}
\end{table}   

Combining this table with the attributes distribution for these 4 clusters, we can define each cluster by their features as follows:
 
 \textbf{Cluster0: Influencers}
This is the smallest and most distinctive cluster. Users belonging to this cluster have a large number of friends and followers, and are included in many lists. In contrast, they have a small number of favourites and statuses. Moreover, their account creation years tend to skew towards the earlier years of the platform, with all accounts being created no later than 2013. Furthermore, all users within this cluster are verified, a feature that is notably less prevalent in other clusters, and have altered the theme or background of their user profile.

 \textbf{Cluster1: Standard users}
This is the largest cluster. These users tend to have a low number of friends, followers, favourites, statuses, and they are not included in many lists. Additionally, their account creation years tend to skew towards the later years of the platform. A notable proportion of users within this cluster have not customized the theme or background of their user profile, and they have the lowest number of users specifying a location or activating geo-location. 

 \textbf{Cluster2: Sharers}
This is the second largest cluster. These users are characterized by having a large amount of favourites and statuses, reflecting the high activity level on the platform and indicating how frequently they engage with others or express their thoughts and opinions.

 \textbf{Cluster3: Lurkers}
This is the second smallest cluster. These users are characterized by having a moderate amount of favourites, while they have a low value of status. We can conclude that they spend time on social media platforms but do not actively contribute content themselves. They might enjoy browsing or engaging with posts that catch their interest, but not actively seeking attention from others.

\subsection{Step 3: Feature Analysis}

\begin{figure*}[h]
    \centering
    \begin{minipage}{0.9\linewidth}
    {\footnotesize
      \includegraphics[width=10cm]{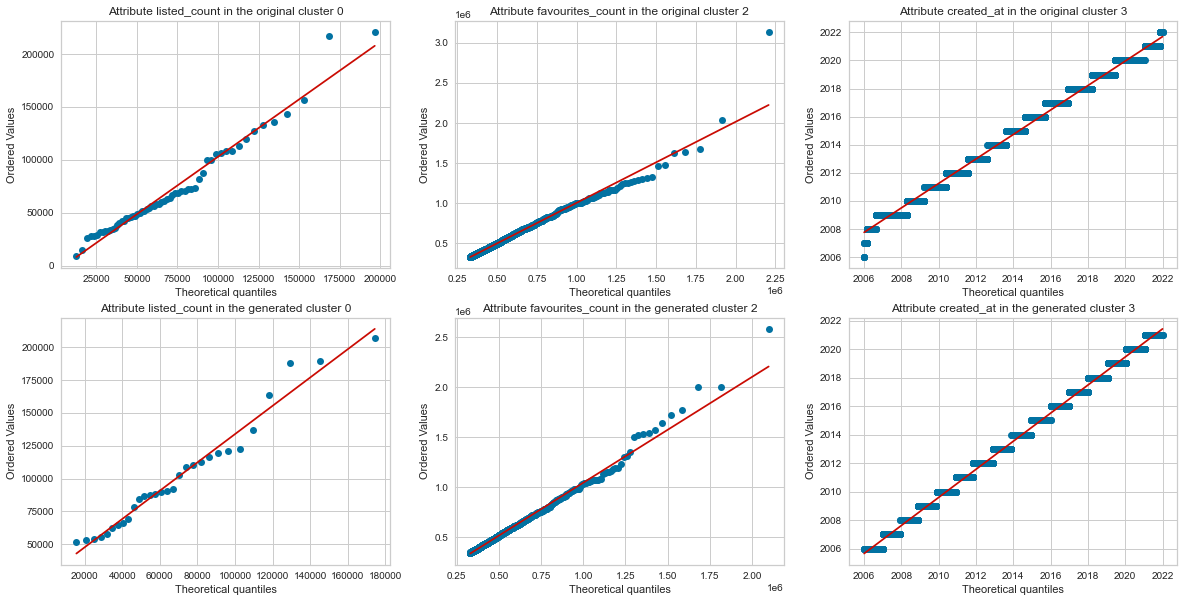}
   
    }
    \end{minipage}
    \caption{Probability plot of the original and generated users}
    \label{fig:probplot}
    \end{figure*}

After the clustering, we assessed the goodness of fit for each distribution using probability plots. The close alignment between the points and the line suggests that the data closely resembles the theoretical distribution. While the majority of distributions exhibited strong fits, certain attributes did not align well with any particular distribution. As a prospective avenue for improvement, we envisage exploring the potential of fitting these attributes to a composite distribution model, aiming to enhance accuracy in future analyses. The first row of Fig.\ref{fig:probplot} shows an example of different attributes of different clusters we got, where data points are plotted against the theoretical quantile line. 

Upon obtaining the distributions, we constructed a model encompassing them and employed this model to compare the distributions with data from another dataset obtained using the same methodology as the original dataset but comprising different users and tweets. This new dataset underwent clustering via K-means. Comparing the model to the new data revealed a notably close fit.

Following this validation, we generated new users based on this model. The number of generated users matched that of the original dataset, with the same percentage of users in each cluster. We also compared these generated users, as the examples showed in the second row of Fig.\ref{fig:probplot}. We provided probability plots for the generated users with matching attributes and matching clusters as the original users depicted. The comparison revealed striking similarities when compared with the clusters from our original dataset, confirming the accuracy of the identified distributions. 

\section{Conclusions}

In this work, we created a social media users database and clustered it using ML technique. We then identified attribute distributions within each cluster to build a model that can generate new users with the characteristics of each cluster.
In summary, our efforts lead to identify four distinct types of Twitter users, each with unique attributes. This allowed us to develop a model capable of generating new users with specified characteristics.
The result we got from this work can assist in further research by developing an agent-based model to simulate user behaviors and interactions.
\begin{credits}
\subsubsection{\ackname} This work has been granted by the Ministerio de Ciencia e Innovación MCIN AEI/10.13039/501100011033 under contract PID2020-113614RB-C21 and by the Generalitat de Catalunya project 2021 SGR 00574.
\end{credits}
%
%
%
\vspace{-1\baselineskip}

\end{document}